\documentstyle[12pt,epsfig]{article}
\def\lsim{\raise0.3ex\hbox{$<$\kern-0.75em\raise-1.1ex\hbox{$\sim$}}}
\def\gsim{\raise0.3ex\hbox{$>$\kern-0.75em\raise-1.1ex\hbox{$\sim$}}}

\def\beq{\begin{equation}}
\def\eeq{\end{equation}}
\def\bea{\begin{eqnarray}}
\def\eea{\end{eqnarray}}
\def\bq{\begin{quote}}
\def\eq{\end{quote}}

\def\gappeq{\mathrel{\rlap {\raise.5ex\hbox{$>$}}
{\lower.5ex\hbox{$\sim$}}}}

\def\lappeq{\mathrel{\rlap{\raise.5ex\hbox{$<$}}
{\lower.5ex\hbox{$\sim$}}}}

\def\Toprel#1\over#2{\mathrel{\mathop{#2}\limits^{#1}}}

\newcommand{\comment}[1]{{}}

\begin{document}

\pagestyle{empty}
\begin{flushright}
HIP-2002-03/TH\\
CERN-TH/2002-007\\
hep-ph/0201256\\
January 2002
\end{flushright}
\vspace*{5mm}
\begin{center}

{\bf CONSTRAINTS FOR NUCLEAR GLUON SHADOWING FROM DIS DATA}\\

\vspace*{1cm}
 K.J. Eskola$^{\rm a,b,}$\footnote{kari.eskola@phys.jyu.fi},
 H. Honkanen$^{\rm a,b}$\footnote{heli.honkanen@phys.jyu.fi}
 V.J. Kolhinen$^{\rm a,b}$\footnote{vesa.kolhinen@phys.jyu.fi} and
 C.A. Salgado$^{\rm c}$\footnote{carlos.salgado@cern.ch}

\vspace{0.3cm}
{\em $^{\rm a}$ Department of Physics, University of Jyv\"askyl\"a,\\
P.O.Box 35, FIN-40351
Jyv\"askyl\"a, Finland\\}
\vspace{0.3cm}
{\em $^{\rm b}$ Helsinki Institute of Physics,\\
P.O.Box 64, FIN-00014 University of Helsinki, Finland\\}
\vspace{0.3cm}
{\em $^{\rm c}$ CERN, Theory Division, CH-1211 Geneva, Switzerland}
\vspace*{2cm}\\
{\bf ABSTRACT} \\ \end{center}
\vspace*{5mm}
\noindent

The $Q^2$ dependence of the ratios of the cross sections of deep
inelastic lepton--nucleus scattering is studied in the framework of
leading twist, lowest order perturbative QCD. The $\log Q^2$ slope of
the ratio $F_2^{\rm Sn}/F_2^{\rm C}$ is computed by using the DGLAP
evolution equations, and shown to be sensitive to the nuclear gluon
distribution functions.  Four different parametrizations for the
nuclear effects of parton distributions are studied. We show that the
NMC data on the $Q^2$ dependence of $F_2^{\rm Sn}/F_2^{\rm C}$ rule
out the case where nuclear shadowing (suppression) of gluons at $x\sim
0.01$ is much larger than the shadowing observed in the ratio
$F_2^A/F_2^{\rm D}$. We also show that the possible nonlinear
correction terms due to gluon fusion in the evolution equations do not
change this conclusion. Some consequences for computation of RHIC
multiplicities, which probe the region $x\gsim0.01$, are also
discussed.

\vfill\eject

\setcounter{page}{1}
\pagestyle{plain}

\section{Introduction}

Nuclear parton distributions (nPDF) are needed in the computation of
inclusive cross sections of hard, factorizable, processes in high
energy nuclear collisions. In the framework of collinear factorization
and leading twist, it is possible to extract universal nuclear parton
distributions $f_{i/A}(x,Q^2)$ from the measurements of deeply
inelastic lepton-nucleus scattering (DIS) and hard processes in $pA$
collisions such as the Drell-Yan process.  The power corrections in
the cross sections \cite{QIU_ht} and in the evolution equations
\cite{GLR,MQ} can be neglected if the scales $Q^2$ and momentum
fractions $x$ involved are large enough. The dependence of the nPDF on
the scale $Q^2$ is then given by the DGLAP evolution equations
\cite{DGLAP}.  Analogously to the global analyses of the parton
distributions of the free proton, the nPDF can be determined based on 
the DGLAP evolution, sum rules and fits to the data. Sets of nPDF
like EKS98  \cite{EKR,EKS} (the code can be found in \cite{EKS98_code,PDFLIB})
and HKM \cite{KUMANO} (Hirai, Kumano and Miyama, the code in \cite{HKM_code})
have become available.

In the DGLAP analyses of parton distributions, the problem boils down
to fixing the initial distributions for the DGLAP evolution at an
initial scale $Q_0^2$.  For the nPDF in particular, there are some
uncertainties in the initial distributions due to the lack of
experimental data. For instance, in the EMC region ($x\gsim 0.2$),
gluons and sea quarks are so far not well constrained \cite{EKST} (see
also \cite{EHKRS,KUMANO} for more discussion). Also, it seems to be a
common belief that the nuclear gluon distributions are constrained
only very weakly by the DIS data at small values of $x$.  In this
letter our aim is to emphasize that this is not the case but very
valuable constraints for the nuclear gluon distributions in the
shadowing region ($x\sim 0.02$) can be obtained from the
$Q^2$ dependence of the structure function ratio $F_2^{\rm
Sn}/F_2^{\rm C}$ measured by the New Muon Collaboration (NMC)
\cite{NMC}.  As first discussed in \cite{PIRNER,EKR}, this is a
consequence of the fact that in the lowest-order DGLAP evolution
$\partial F_2(x,Q^2)/\partial\log Q^2\sim \alpha_s xg(2x,Q^2)$
\cite{PRYTZ} at small $x$.  One of the tasks in the present paper is
to compare the results of the HKM analysis \cite{KUMANO}, where the
constraint from the measured $Q^2$ dependence of $F_2^{\rm
Sn}/F_2^{\rm C}$ has not been applied, with the results obtained with
EKS98 \cite{EKR,EKS}, which makes use of this constraint.

The Relativistic Heavy Ion Collider (RHIC) initiated the collider era
for the search of the Quark-Gluon Plasma in ultrarelativistic $AA$
collisions. From the point of view of the nPDF, the measurements of
charged particle multiplicities $dN_{\rm ch}/d\eta$ in Au+Au
collisions \cite{PHOBOS,PHENIX,STAR,BRAHMS}
have been truly exciting, since production of semihard
gluons at 1...2 GeV scales is expected to dominate particle and
entropy production at $\sqrt s \gsim 200$ GeV.  The measured
multiplicities are thus a probe of the nuclear gluon distributions at
$x\sim 0.01$ at these scales: in the models employing saturation
\cite{GLR,BM,McLV,EKRT,DK}, the multiplicity scales as $dN_{\rm
ch}/d\eta\sim xg_A(x_{\rm sat},Q_{\rm sat}^2)$ (with possible powers
of $\alpha_s$ added). In the two-component (hard+soft) models, such as
HIJING \cite{HIJING,newHIJING}, the nPDF are probed through the
perturbative (hard) minijet component.

Recently, in \cite{newHIJING}, it has been suggested based on the
HIJING model that the RHIC data on multiplicities would indicate that
the gluons were more strongly shadowed than the sea quarks. To challenge
this interesting suggestion, we shall perform a DGLAP analysis of the
nPDF based on the initial nuclear effects for quark and gluon
distributions as given in \cite{newHIJING}. As the second and main
point of this paper, we shall show that in the leading twist, lowest
order DGLAP framework the NMC data on the $Q^2$ dependence of
$F_2^{\rm Sn}/F_2^{\rm C}$ \cite{NMC} rules out very strongly shadowed
gluons. In relation to HIJING, consequences of this observation 
for the parameter $p_0$, which determines the
division into soft and hard components, will be discussed.

When the density of gluons in the wave functions of the colliding
nuclei (or hadrons) becomes large enough, gluon fusion starts to play
a role.  This induces non-linearities into the QCD evolution equations
\cite{GLR,MQ,JALILIAN,KOVCHEGOV,ARMESTO} at small values of $x$ and
$Q^2$. As the last subject to study in this letter, we add the
non-linear terms (GLRMQ) into the DGLAP equations for Sn and C nuclei.
Comparison with the NMC data \cite{NMC} shows, that a very strong
gluon shadowing is ruled out even more clearly when the
non-linearities are included.

\section{DGLAP analysis and different parametrizations}

Following the notation in our previous works \cite{EKR,EKS,EHKRS}, we
define the nPDF through the nuclear effects,
\begin{equation}
  R_{i}^A(x,Q^2)\equiv{f_i^A(x,Q^2)\over f_i(x,Q^2)}, \ \ \ \ \ 
  i=g,\ u,\ d,\ s,\ \bar u,\ \bar d,\ \bar s, \dots ,
\label{eq1}
\end{equation} 
where $f_i^A\equiv f_i^{p/A}$ is the number density distribution of a
flavour $i$ in a bound proton of a nucleus $A$, and $f_i$ is the
corresponding distribution in the free proton.  The parton
distributions of bound neutrons in isoscalar nuclei are obtained
through isospin symmetry, $f_{u(\bar u)}^{n/A}=f_{d(\bar d)}^{p/A}$
and $f_{d(\bar d)}^{n/A}=f_{u(\bar u)}^{p/A}$, and we expect this to
be a good approximation for non-isoscalar nuclei as well.  Nuclear
effects in the structure function $F_2^A$ we define through
\begin{eqnarray}
R_{F_2}^A(x,Q^2) 
= \frac{\frac{1}{A}F_2^A}{\frac{1}{2}F_2^{\rm D}}
\approx \frac{\frac{1}{2}(F_2^{p/A} + F_2^{n/A})
+\frac{1}{2}(\frac{2Z}{A}-1)(F_2^{p/A} - F_2^{n/A})}
      {\frac{1}{2}(F_2^{p} + F_2^{n})},
\label{RF2}
\end{eqnarray}
where $F_2$ of the protons (neutrons) of the nucleus $A$ is
$F_2^{p(n)/A}(x,Q^2)= \sum_{q} e_q^2[xf_q^{p(n)/A}(x,Q^2)+xf_{\bar
q}^{p(n)/A}(x,Q^2)]$, and where the small nuclear effects in D have
been neglected.

In order to explicitly show the effect of nuclear gluon shadowing to the
scale dependence of the ratio $F_2^{\rm Sn}/F_2^{\rm C}$, we 
shall study different parametrizations for the nuclear effects. 
On one hand, we will directly make use of the available results from 
the global DGLAP analyses of nPDF, EKS98 \cite{EKR,EKS} and HKM \cite{KUMANO}. 
On the other hand, we perform the DGLAP evolution for nPDF by taking the 
initial modifications $R_i^A(x,Q_0^2)$ from two different parametrizations
in which the $Q^2$ dependence is assumed to be negligible:
\begin{itemize}
\item 
HPC parametrization \cite{HPC} which is based on \cite{VARY}:
\begin{eqnarray}
\nonumber
  R_i^A(x)=R_{F_2}^A(x) &=& 
     R_{\rm sh} \frac{\displaystyle 1+c_Dc_A(1/x-1/x_{\rm sh})}
                    {\displaystyle 1+c_AA^{p_A}(1/x-1/x_{\rm sh})}
      \Theta(x_{sh}-x)+ \\
  &&\hspace{-1cm}+(a_{\rm emc}-b_{\rm emc}x) \Theta(x-x_{sh})\Theta(x_f-x)+
  R_f\bigg(\frac{\displaystyle 1-x_f}{\displaystyle 1-x}\bigg)^{p_{\rm f}}
  \Theta(x-x_f)
\label{HPC_fit}
\end{eqnarray}
where the different regions are matched together by setting
$R_{\rm sh}=a_{\rm emc}-b_{\rm emc}x_{\rm sh}$, 
$R_f = a_{\rm emc}-b_{\rm emc}x_f$ and 
$a_{\rm emc} = 1 + b_{\rm emc}x_{\rm emc}$. 
The $A$ dependence of $b_{\rm emc}$ is $b_{\rm emc}= p_{\rm emc}[
1-A^{-1/3}-1.145 A^{-2/3}+0.93 A^{-1} + 0.88 A^{-4/3}-0.59 A^{-5/3}]$ 
from Ref. \cite{SMIRNOV}.  A fit to DIS data results in 
$p_A= 0.10011$, $c_A=0.0127343$, $c_D = 1.05570$ $x_{\rm sh}  = 0.154037$,
$x_{\rm emc} = 0.275097$, $p_{\rm emc} = 0.525080$,
$x_{\rm f} = 0.742059$, and $p_{\rm f} = 0.320992$.

\item
New HIJING parametrization \cite{newHIJING} which replaces the
obsolete one in HIJING \cite{HIJING} (see also the discussion in
\cite{EHKRS}). In the new parametrization, the quark sector is fitted
to modern DIS data, while the gluon sector, especially shadowing and
antishadowing are constrained by the requirement that (with the
updated $\sqrt s$ dependence of the cut-off parameter $p_0$ for the
transverse momentum of minijets) HIJING reproduces the measured
charged-particle multiplicities in Au+Au collisions at RHIC. In this 
way, gluon shadowing is suggested to be much stronger than that of
quarks:
\begin{eqnarray}
  R_q^A(x)=R_{\bar q}^A=R_{F_2}^A(x)
&=&1.0+1.19\log^{1/6}\!A\;(x^3-1.2x^2+0.21x) 
\nonumber \\
 &&-s_q\;(A^{1/3}-1)^{0.6}(1-3.5\sqrt{x})\exp(-x^2/0.01)\label{HIJING_fitq}\\
\nonumber
\\
  R_g^A(x)&=&1.0+1.19\log^{1/6}\!A\;(x^3-1.2x^2+0.21x) 
\nonumber \\
        &&-s_g\;(A^{1/3}-1)^{0.6}(1-1.5x^{0.35})\exp(-x^2/0.004),
\label{HIJING_fitg}
\end{eqnarray}

\noindent
with $s_q=$ 0.1 and $s_g$= 0.24\dots 0.28. Below, we shall use the value 0.24.
\end{itemize} 

\begin{figure}[tb]
\vspace{-0.5cm}
\centering{\epsfxsize=10cm\epsfbox{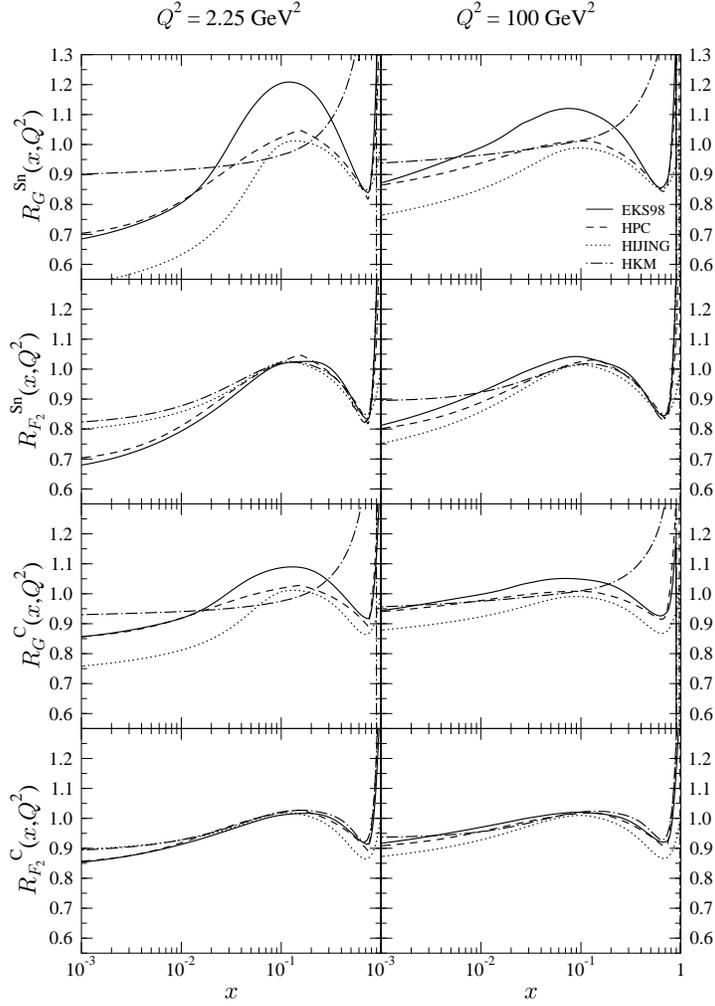}}
\vspace{0cm}
\caption[a]{{\small $R_g(x,Q^2)$ and $R_{F_2}(x,Q^2)$ for Sn ($A=117$)
and C ($A=12$) as function of $x$ for scales $Q^2=Q_0^2=2.25$ GeV$^2$ and
$Q^2=100$ GeV$^2$. The difference of the ratios 
$R_g^A(2x,Q^2)/R^A_{F_2}(x,Q^2)$ for Sn and C is proportional to the
slope $\partial(F_2^{\rm Sn}/F_2^{\rm C})/\partial\log Q^2$, as shown
in Eq. (\ref{RF2SnCslope}). } }
\vspace{0cm}
\label{FigRgRF2}
\end{figure}

In Fig. \ref{FigRgRF2}, in the left panels, we plot the nuclear
effects for gluons in tin ($A=117$, top row) and carbon ($A=12$, third
row) nuclei at a scale $Q_0^2=2.25$ GeV$^2$, as given by EKS98 and
HKM, and by the HIJING and HPC parametrizations above.  For further
discussion, we also show $R_{F_2}^A(x,Q_0^2)$ from Eq. (\ref{RF2}) (2nd
and 4th rows, correspondingly). With the EKS98, HIJING and HPC
parametrizations, we have used the MRST (central gluon) LO PDFs of the
free proton \cite{MRST,PDFLIB}. The HKM results are obtained directly from
the HKM code \cite{HKM_code}, based on \cite{KUMANO} and where 
the LO MRST (c-g) PDFs are also used.

As seen in Fig. \ref{FigRgRF2}, for the gluons there are quite
distinctive differences between the sets used: in EKS98 there is
strong gluon antishadowing, which originates from the requirement of
conservation of momentum combined with the constraint obtained from
the measured $Q^2$ dependence of $F_2^{\rm Sn}/F_2^{\rm C}$.  
In HKM in turn, the gluons are less shadowed at small values of $x$ but no
EMC-effect appears at $x\sim 0.3\dots0.7$. Momentum is conserved also in HKM, 
so the deficit of momentum at $x\lsim0.2$ is compensated by a rapid
increase of $R_g^A$ at $x\gsim 0.2$. Of the four cases studied here,
the HIJING parametrization has the strongest gluon shadowing. Since no
antishadowing appears, the HIJING parametrization underestimates the
momentum sum at $Q_0^2$ by about 10~\%. As discussed in \cite{EHKRS}, 
the HPC parametrization underestimates the momentum sum at $Q_0^2$ by 5~\%. 

\begin{figure}[tb]
\vspace{-0.5cm}
\centering{\epsfxsize=15cm\epsfbox{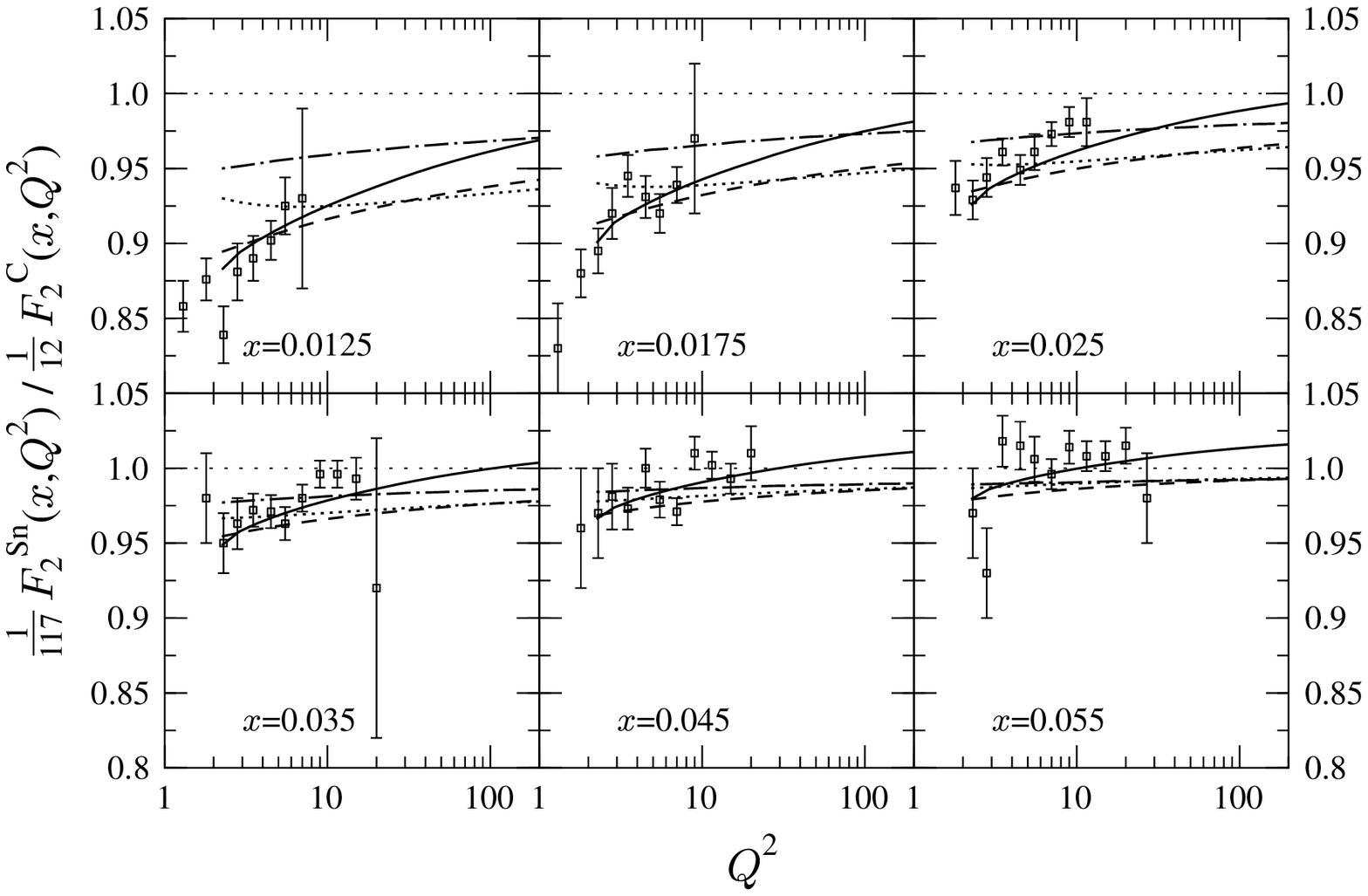}}
\vspace{-1cm}
\caption[a]{{ \small Comparison of the calculated and measured $Q^2$
dependence of the ratio $F_2^{\rm Sn}/F_2^{\rm C}$. The NMC data
\cite{NMC} are shown with statistical errors only. The results for
EKS98 \cite{EKR,EKS} (solid lines) and HKM \cite{KUMANO}
(dotted-dashed) are from the corresponding global DGLAP analyses. The
$Q^2$ dependence of the  HPC (dashed) and HIJING (dotted) cases is obtained
from the DGLAP equations by taking the initial conditions
for the nuclear effects from Eqs. (\ref{HPC_fit}) and
(\ref{HIJING_fitq})-(\ref{HIJING_fitg}).
}}
\vspace{0cm}
\label{FigRF2}
\end{figure}

Regarding the ratio $R_{F_2}^A$, all parametrizations are quite
similar at $x\gsim 0.02$, as they are based on the fits to the DIS
data.  At $x\lsim 0.005$, however, due to the lack
of DIS data in the perturbative region, some differences arise.  Note
also that in Fig. \ref{FigRgRF2} the ratio $R_{F_2}^A$ from EKS98 is
plotted for isospin symmetrized $A=117$ (in order to compare with the
NMC data \cite{NMC}) but the HKM-results are for non-isospin symmetric
tin, and with the small effects for $D$ included, as obtained directly
from the numerical code \cite{HKM_code} by HKM.  At the region of small 
values of $x$, where the focus of the present paper is, the isospin 
effects can in any case be safely neglected.

Next, we consider the (lowest order) DGLAP evolution of the nPDF and
explicitly show the consequences of different assumptions of gluon
shadowing. In EKS98 and HKM the scale evolution has already been done
but for the other two cases it needs to be performed separately.  For
the HPC and HIJING nuclear effects, we do this by choosing $Q_0^2$ as
the initial scale, and computing the intial distributions with the
MRST distributions, taking the initial nuclear effects from
Eqs. (\ref{HPC_fit}) and (\ref{HIJING_fitq})-(\ref{HIJING_fitg}),
correspondingly.  The absolute nPDF are then evolved from $Q_0^2$ to
higher scales with the DGLAP equations. The results for $R_G^A(x,Q^2)$
and $R_{F_2}^A(x,Q^2)$ at a scale $Q^2=100$ GeV$^2$ from all cases
studied, are shown in the right panels of Fig.  \ref{FigRgRF2}. Again,
with the EKS98 we use the MRST distributions in plotting the ratio
$R_{F_2}^A$, and the HKM results are directly from the HKM code
\cite{HKM_code}.

Two observations can be immediately made: first, the pQCD scale
evolution is a sizable effect for the gluon ratios at $x\lsim 0.3$,
especially in the small-$x$ region. Second, at small values of $x$,
due to the very strong gluon shadowing, the HIJING parametrization 
predicts the slope $\partial R_{F_2}^A(x,Q^2)/\partial \log Q^2$ 
to be {\em negative}, contrary to the other three cases studied. 

The comparison between the calculated results and the NMC data
\cite{NMC} for the ratio $F_2^{\rm Sn}/F_2^{\rm C}$ is shown in
Fig.~\ref{FigRF2}.  The EKS98 parametrization reproduces well the
experimental data, the $\log Q^2$ slopes in particular, which is not
surprising as these data have been taken into account in the analysis
\cite{EKR,EKS}.  The DGLAP-evolved HPC also reproduces the data
reasonably well. The HKM results miss the absolute normalization of
the data at the smallest values of $x$ but have the right kind of
curvature.  The data would also suggest faster evolution at small
scales than that in HKM. The HIJING parametrization results
in a negative $Q^2$-slope for the ratio $F_2^{\rm Sn}/F_2^{\rm C}$ at
small values of $x$ and $Q^2$. This clearly is in contradiction with the
data. This behaviour is caused by the strong gluon shadowing in the 
HIJING parametrization at small $x$, as will be discussed next.

At small values of $x$, the structure function $F_2$ (of both $p$ and $n$)
is dominated by  sea quarks, whose DGLAP evolution in turn is dominated 
by gluons. The $\log Q^2$ slope of $F_2$ can be approximated at lowest order 
as \cite{PRYTZ}
\begin{equation}
\frac{\partial F_2^{p(n)}(x,Q^2)}{\partial \log Q^2} \approx
\frac{10\alpha_s}{27\pi} xg(2x,Q^2).
\end{equation}
This leads to
\begin{equation} 
\frac{\partial R_{F_2}^A(x,Q^2)}{\partial \log Q^2}
\approx
\frac{10\alpha_s}{27\pi}\frac{xg(2x,Q^2)}{\frac{1}{2}F_2^{\rm D}(x,Q^2)}
\biggl\{R_g^A(2x,Q^2)-R_{F_2}^A(x,Q^2)\biggr\},
\label{RF2slope}
\end{equation}
and 
\begin{eqnarray}
  \frac{\partial (\frac{1}{117}F_2^{\rm Sn}/\frac{1}{12}F_2^{\rm C})}
       {\partial \log Q^2}
 & \approx &\frac{10 \alpha_s}{27\pi}\frac{x g(2x,Q^2)}{\frac{1}{2}F_2^{\rm D}(x,Q^2)}
    \frac{R_{F_2}^{\rm Sn}(x,Q^2)}{R_{F_2}^{\rm C}(x,Q^2)} \left\{ 
    \frac{R_{g}^{\rm Sn }(2x,Q^2)}{R_{F_2}^{\rm Sn}(x,Q^2)} -
    \frac{R_{g}^{\rm C }(2x,Q^2)}{R_{F_2}^{\rm C}(x,Q^2)} \right\},
    \label{RF2SnCslope}
\end{eqnarray}
where $xg$ is the gluon distribution in the free proton, and 
$F_2^{\rm D}=F_2^p+F_2^n$.

In the HIJING parametrization gluons are much more strongly shadowed
than quarks, so $R_g^A(2x,Q_0^2)<
R_q^A(x,Q_0^2)$. Eq. (\ref{RF2slope}) thus directly shows why a
negative $\log Q^2$ slope for the ratio $R_{F_2}^A$ seen in
Fig. \ref{FigRgRF2} is bound to follow.  Based on
Eq. (\ref{RF2SnCslope}) we can also understand the origin for the
differences between the computed $\log Q^2$ slopes of the ratio
$F_2^{\rm Sn}/F_2^{\rm C}$.  The second term in the curly brackets in
Eq. (\ref{RF2SnCslope}) is obviously always closer to unity than the
first one, since the nuclear effects in smaller nuclei are smaller.
On the other hand, the NMC data in Fig. \ref{FigRF2} indicates a
clearly positive $\log Q^2$-slope.  These facts imply that $R_g^A$ is
bounded from below as $R_g^A(2x,Q^2)>R_{F_2}^{\rm A}(x,Q^2)$.
Consequently, $R_g^A(x,Q_0^2)$ cannot be much smaller than
$R_{F_2}^{\rm A}(x,Q_0^2)$.  The values of $R_g^A(2x,Q_0^2)$ and
$R_{F_2}^{\rm A}(x,Q_0^2)$ at $x=0.0125$ for Sn and C are directly
readable off from Fig. \ref{FigRgRF2}. Both the sign and the relative
order in the magnitude of the computed slopes in the first panel of
Fig. \ref{FigRF2} can be understood by substituting these values in
Eq. (\ref{RF2SnCslope}).  The fact that the $\log Q^2$ slopes from the
HIJING parametrization of nuclear effects are opposite to the measured
ones leads us to the conclusion that very strongly shadowed gluons are
ruled out within the leading twist DGLAP framework.

The observation above has an interesting consequence for the HIJING
model.  As discussed in \cite{newHIJING}, particle production in $AA$
collisions in HIJING is due to contributions from soft and hard
components,
\begin{equation}
\frac{dN_{ch}^{AA}}{d\eta}= \frac{1}{2}\langle N_{\rm parts}^{AA}\rangle
\langle n_{\rm soft}\rangle + \langle N_{\rm binary}^{AA}\rangle 
\langle n_{\rm hard}\rangle\frac{\sigma_{\rm
jet}^{AA}(\sqrt s, p_0)}{\sigma_{\rm in}^{NN}(\sqrt s)},
\label{Nch_HIJING}
\end{equation}
where $\langle N_{\rm parts}\rangle \sim 2A$ is the average number of
participants in $AA$ for the centrality selection considered, $\langle
n_{\rm soft}\rangle=1.6$ is the average multiplicity from soft
processes to the rapidity interval considered, $\langle N_{\rm
binary}\rangle \sim A^{4/3}$ is the average number of binary
nucleon-nucleon collisions, $\langle n_{\rm hard}\rangle=2.2$ represents
particle production from (mini)jet hadronization, and $\sigma_{\rm
in}^{NN}(\sqrt s) $ is the inelastic nucleon-nucleon cross
section. The integrated minijet cross section $\sigma_{\rm
jet}^{AA}(\sqrt s, p_0)$ is computed perturbatively by using the nPDF.
The parameter $p_0$ for the minimum transverse momentum of minijets defines
the division into soft and hard components. The values of $p_0$ are
determined based on fits to the inelastic cross sections measured in $pp$ and
$p\bar p$ collisions. The $\sqrt s$ dependence of $p_0$ in HIJING has now
been updated \cite{newHIJING}, and $p_0$ is an increasing function of 
$\sqrt s$. Unlike the saturation scale in models with parton saturation 
\cite{GLR,BM,McLV,EKRT,DK}, the scale $p_0$ does not depend on $A$.

We write $\sigma_{\rm jet}^{AA}(\sqrt s, p_0) = [R_g^A(\langle
x\rangle,\langle Q^2\rangle)]^p\sigma_{\rm jet}(\sqrt s,p_0)$, where
$\langle x\rangle\approx 2p_0/\sqrt s$ and $\langle Q^2\rangle=
ap_0^2$ with $a\sim 2$ for the minijet production at central
rapidities.  The effective power $p=1\dots2$ describes the net effect
of gluon shadowing. From Eq. (\ref{Nch_HIJING}) we then obtain
\begin{equation}
[R_g^A(\frac{2p_0}{\sqrt s},ap_0^2)]^p
\sigma_{\rm jet}(\sqrt s,p_0) = \frac{\sigma_{\rm
in}^{NN}(\sqrt s)}{\langle N_{\rm binary}^{AA}\rangle \langle n_{\rm
hard}\rangle}\left[\frac{dN_{ch}^{AA}}{d\eta}-\frac{1}{2}\langle N_{\rm
parts}^{AA}\rangle \langle n_{\rm soft}\rangle\right].
\label{sigmaAA}
\end{equation}
Requiring that the model reproduces the measured multiplicity in Au+Au
collisions at RHIC, the r.h.s. of Eq. (\ref{sigmaAA}) is a fixed number
for each $\sqrt s$ and $A$, and thus independent of $p_0$. 
As discussed above, within the DGLAP framework the NMC data implies
that $R_g^A$ should be larger than that of the HIJING
parametrization in Eq. (\ref{HIJING_fitg}) (this is due to both weaker 
gluon shadowing and due to scale evolution). Correspondingly, $\sigma_{\rm
jet}(\sqrt s, p_0)$ should be smaller. This in turn implies that $p_0$
should be larger than in the $pp$ case.  We are thus lead to the
conclusion that the same multiplicities as are currently obtained from the
HIJING model with an $A$-independent $p_0(\sqrt s)$ and very strong gluon 
shadowing, can be obtained with weaker gluon shadowing by introducing a 
scale $p_0$ which is an increasing function of both $\sqrt s$ and $A$.

\section{Non-linear effects in $Q^2$ evolution}

The DGLAP scale evolution discussed above is linear in the parton
densities.  At very small values of $x$ the density of gluons
increases to the extent that contributions from non-linear phenomena
due to gluon fusion may start to play a role. The first non-linear
correction terms to pQCD evolution equations have been computed in
\cite{GLR,MQ}, let us call them ``GLRMQ terms''. One could argue that
perhaps these corrections would change the scale evolution in such a
way that stronger gluon shadowing could be allowed. In order to study
this possibility, we next add the GLRMQ terms into the DGLAP
equations. In connection with the DGLAP analysis of nPDF, the non-linear 
effects have been numerically studied e.g. in \cite{QIU} (see also
\cite{OTHERS}). 

The valence quark evolution remains unmodified, but for the gluons and
sea quarks the generic form of the equations reads
\begin{eqnarray}
\frac{\partial xg_A}{\partial\log Q^2} &=& 
\bigg(\frac{\partial xg_A}{\partial\log Q^2}\bigg)_{\rm DGLAP} 
- \frac{1}{Q^2}{\cal R}_{ggg}(x,Q^2) 
\label{MQeqs_g}\\
\frac{\partial x\bar q_A}{\partial\log Q^2} &=& 
\bigg(\frac{\partial x\bar q_A}{\partial\log Q^2}\bigg)_{\rm DGLAP} - 
\frac{1}{Q^2}\bigg[{\cal R}_{\bar qgg}(x,Q^2)-
{\cal R}_{\bar qgg}^{\rm HT}(x,Q^2) \bigg]
\label{MQeqs_q}
\\
\frac{\partial x g_{HT}}{\partial\log Q^2} &=& 
 - {\cal R}_{ggg}(x,Q^2)
\label{MQeqs_ht}
\end{eqnarray}
where GLRMQ terms ${\cal R}_{ggg}\sim \alpha_s^2\int_x^1 (dy/y)
y^2g_A^{(2)}(y,Q_0^2)$, ${\cal R}_{\bar qgg}\sim \alpha_s^2
x^2g_A^{(2)}(x,Q_0^2)$ and ${\cal R}_{\bar qgg}^{\rm HT}\sim
\alpha_s\int_x^1(dy/y)(x/y)\bar \gamma_{\rm FG}(x/y)yg_{\rm
HT}(y,Q^2)$. The detailed form of these terms can be found in
\cite{MQ}.  Following \cite{MQ} and \cite{QIU}, for the 2-gluon
density we take $x^2g_A^{(2)}(x,Q^2)= \frac{A}{\pi
R_A^2}[xg_A(x,Q^2)]^2$, where $\frac{A}{\pi R_A^2}$ and the nuclear
radius $R_A=1.12A^{1/3}-0.86A^{-1/3}$ are based on the Woods-Saxon
parametrization of nuclear densities. For the gluon higher-twist term,
we assume $xg_{HT}(x,Q_0^2)=x^2g_A^{(2)}(x,Q_0^2)$ and that
$g_{HT}(x,Q^2)\ge0$.

\begin{figure}[tb]
\vspace{-0.5cm}
\centerline{
\epsfxsize=15cm\epsfbox{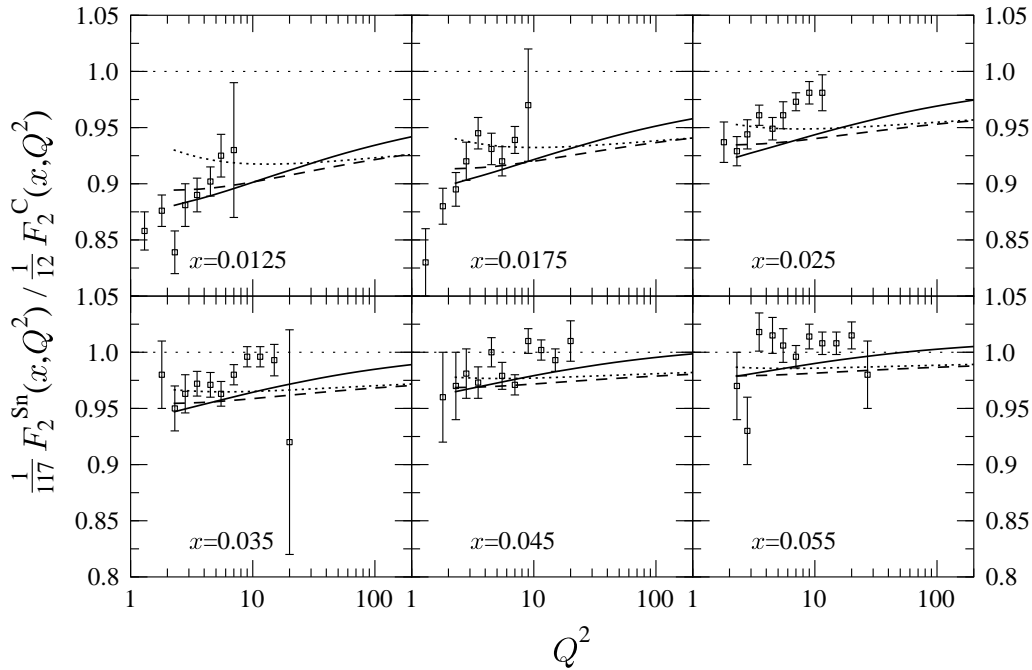}}
\vspace{-1cm}
\caption[a]{{ \small The scale dependence of the ratio $F_2^{\rm
   Sn}/F_2^{\rm C}$ calculated using DGLAP evolution with MQ
   corrections, and compared to NMC data \cite{NMC}. Initial
   conditions for nuclear effects are taken from EKS98 (solid lines),
   HPC (dashed) and HIJING (dotted) parametrizations.   }}
\vspace{-0cm}
\label{FigRF2MQ}
\end{figure}

We obtain the initial conditions for the actual nPDF as before, from
the EKS98, HIJING and HPC parametrizations and with the MRST distributions
for the free proton. We do not make an attempt to include the GLRMQ terms to
the HKM analysis. The results for the scale evolution of the ratio
$\frac{1}{117}F_2^{\rm Sn}/\frac{1}{12}F_2^{\rm C}$ are shown in
Fig.~\ref{FigRF2MQ} against the NMC data. We observe that the effects of 
the GLRMQ corrections remain fairly modest (as they should, in order to 
stay as corrections) but that they make the positive $\log Q^2$-slopes 
(EKS98, HPC) flatter, and the negative slopes (HIJING) even more negative 
than in the case of pure DGLAP evolution. Our conclusion therefore is that 
the GLRMQ terms do not give support to the strong shadowing of gluons, either.

The systematics of the change in the $\log Q^2$ slopes at the initial
scale $Q_0^2$ can also be easily understood from Eqs. (\ref{MQeqs_q}). The
evolution equation for $F_2^A$ can be written as $F_2^{A \prime} =
F_{2\, {\rm DGLAP}}^{A\, \prime} + F_{2\, {\rm GLRMQ}}^{A\, \prime},$
where the prime stands for $\partial/\partial\log Q^2$, and where the
second term contains only the GLRMQ corrections for the sea quark
evolution from Eq. (\ref{MQeqs_q}).  The net effect of the GLRMQ
corrections to the $\log Q^2$ slope of sea quarks is negative, and
dominated by the term ${\cal R}_{\bar qgg}\sim
A^{1/3}(R_g^A)^2(xg)^2$. The $\log Q^2$ slope of the the ratio
$F_2^{\rm Sn}/F_2^{\rm C}$ can then be expressed as
\begin{equation} 
  \frac{\partial (F_2^{\rm Sn}/F_2^{\rm C})}{\partial \log Q^2} 
  = \frac{F_2^{\rm Sn}}{F_2^{\rm C}} \left\{
    \bigg[ \frac{F_{2\,{\rm DGLAP}}^{\rm Sn \, \prime}}{F_2^{\rm Sn}} 
          - \frac{F_{2\,{\rm DGLAP}}^{\rm C \, \prime}}{F_2^{\rm C }} \bigg] 
  + 
    \bigg[ \frac{F_{2\,{\rm GLRMQ}}^{\rm Sn \,\prime}}{F_2^{\rm Sn}} 
          - \frac{F_{2\,{\rm GLRMQ}}^{\rm C \, \prime}}{F_2^{\rm C }} \bigg]
    \right\}
\end{equation}
where the latter term in brackets again contains only the GLRMQ contributions
to the $\log Q^2$ slope from Eq. (\ref{MQeqs_q}). Using 
$F_2^A\sim R_{F_2}^A(F_2^p+F_2^n)$, the GLRMQ part becomes
\begin{equation} 
\frac{F_{2\,{\rm GLRMQ}}^{\rm Sn \,\prime}}{F_2^{\rm Sn}} 
          - \frac{F_{2\,{\rm GLRMQ}}^{\rm C \, \prime}}{F_2^{\rm C }}
\sim 
\frac{12^{1/3}(R_g^{\rm C})^2}{R_{F_2}^{\rm C}}
\bigg[
1- \left(\frac{117}{12}\right)^{1/3} \frac{R_{F_2}^{\rm C}}{R_{F_2}^{\rm Sn}}
\left(\frac{R_g^{\rm Sn}}{R_g^{\rm C}}\right)^2
\bigg].
\label{MQapprox}
\end{equation}
Reading the values for $R_{F_2}^A(x,Q_0^2)$ and $R_g^A(x,Q_0^2)$ at
$x\sim0.01$ off from Fig. 1, it is easy to see that the contribution
from the GLRMQ terms to the $\log Q^2$ derivative of the ratio
$F_2^{\rm Sn}/F_2^{\rm C}$ is indeed negative in all the three cases
studied.

We emphasize that for the DGLAP+GLRMQ case we have not attempted to
make a global analysis of the nPDF, where the initial conditions would
be based on fits to various set of data. Such an analysis will require also
detailed studies of the constraints for the magnitude of the
nonlinearities in the case the free proton. We note, however, that
according to Fig. \ref{FigRF2MQ}, in the framework of DGLAP+GLRMQ we
could expect somewhat less gluon shadowing for $g_A(x,Q_0^2)$ than in
the framework of pure DGLAP. This will also lead to slightly smaller 
excess of gluons at larger values of $x$ (see EKS98 and HKM in Fig. 1). 
This analysis is left as a future task.

\section{Conclusions}
We have studied the $Q^2$ dependence of the ratio $F_2^{\rm
Sn}/F_2^{\rm C}$ in the pQCD framework of lowest-order DGLAP evolution
and leading twist. We emphasize that the NMC data on the $Q^2$
dependence of $F_2^{\rm Sn}/F_2^{\rm C}$ \cite{NMC} is the most direct
(although indirect) measurement of the nuclear gluon distributions that
is currently available. These data provide a valuable constraint for
pinning down the nuclear gluon shadowing in the global DGLAP analyses
of nPDFs.

We have demonstrated the sensitivity of the $\log Q^2$ slopes of
$F_2^{\rm Sn}/F_2^{\rm C}$ to the gluon shadowing by comparing four
different approaches. The NMC data \cite{NMC} implies that the nuclear
effects in the gluon distributions ($R_g^A$), should lie close to
those in $F_2^A/F_2^{\rm D}$ at $x\sim 0.01$ and $Q^2\sim 2$ GeV$^2$. In
particular, a very strong gluon shadowing, as suggested
e.g. in \cite{newHIJING}, is ruled out by the NMC data, since it leads to 
a $\log Q^2$ slope whose sign is opposite to what is measured.
Consequences for the HIJING model have been discussed in Sec. 2.  We
also suggest to use the NMC data \cite{NMC} and \cite{NMC_AvsC} as
further constraints in the DGLAP analysis of HKM \cite{KUMANO}.

The DGLAP approaches \cite{EKR,EKS} and \cite{KUMANO} conserve momentum
explicitly. The scale independent parametrizations from HIJING
\cite{HIJING,newHIJING} and HPC \cite{HPC} do not do that, 
especially not if gluons are very strongly shadowed and no
antishadowing appears.  We emphasize that the strong antishadowing in
the gluons of EKS98 results from the fact that the gluon shadowing is
constrained by the NMC data at $x\sim 0.01$, and to compensate the
loss of momentum there, antishadowing is needed.

We have also studied the effects of the nonlinear GLRMQ terms
\cite{GLR,MQ} in the DGLAP equations. These terms decrease the slope
$\partial (F_2^{\rm Sn}/F_2^{\rm C})/\partial\log Q^2$. For the very
strongly shadowed gluons this results in even more negative slopes
than without the GLRMQ corrections. Our conclusion therefore is that
within the DGLAP and DGLAP+GLRMQ frameworks studied we cannot find any
support from the DIS data for a much stronger gluon shadowing at 
$x\sim 0.01$ and $Q^2\sim 2$ GeV$^2$ than what is observed in the 
ratio $F_2^A/F_2^{\rm D}$. To resolve the situation at smaller values of 
$x$, especially in the region relevant for the few-GeV scales at the LHC,
more DIS data in the perturbative region would be needed.

\section*{Acknowledgements}
We thank N. Armesto, P.V. Ruuskanen and X.-N. Wang for discussions.
Financial support from the Academy of Finland, grant no. 50338,
is gratefully acknowledged.  C.A.S. is supported by a Marie Curie
Fellowship of the European Community programme TMR (Training and
Mobility of Researchers), under the contract number
HPMF-CT-2000-01025.

\end{document}